\documentclass{ws-procs975x65}

\begin{document}

\title{Baryon-meson interactions in chiral quark model\footnote{\uppercase{T}his work is supported by the \uppercase{N}ational \uppercase{N}atural \uppercase{S}cience \uppercase{F}oundation of \uppercase{C}hina \uppercase{N}o. 10475087.}}

\author{F. Huang, Z. Y. Zhang and Y. W. Yu}

\address{Institute of High Energy Physics, P.O. Box
918-4, Beijing 100049, China}

\maketitle

\abstracts{Using the resonating group method (RGM), we dynamically
study the baryon-meson interactions in chiral quark model. Some
interesting results are obtained: (1) The $\Sigma K$ state has an
attractive interaction, which consequently results in a $\Sigma K$
quasibound state. When the channel coupling of $\Sigma K$ and
$\Lambda K$ is considered, a sharp resonance appears between the
thresholds of these two channels. (2) The interaction of $\Delta
K$ state with isospin $I=1$ is attractive, which can make for a
$\Delta K$ quasibound state. (3) When the coupling to the $\Lambda
K^*$ channel is considered, the $N\phi$ is found to be a
quasibound state in the extended chiral SU(3) quark model with
several MeV binding energy. (4) The calculated $S$-, $P$-, $D$-,
and $F$-wave $KN$ phase shifts achieve a considerable improvement
in not only the signs but also the magnitudes in comparison with
other's previous quark model study.}

\section{Introduction}

Nowadays people still need QCD-inspired models to study the
non-perturbative QCD effects in the low-energy region. Among these
models, the chiral SU(3) quark model has been quite successful in
reproducing the energies of the baryon ground states, the binding
energy of deuteron, the $NN$ scattering phase shifts, and the $NY$
(nucleon-hyperon) cross sections. Inspired by these achievements,
we try to extend this model to study the baryon-meson systems. In
order to study the short-range feature of the quark-quark
interaction in the low-energy region, we further extend our chiral
SU(3) quark model to include the vector meson exchanges. The OGE
that dominantly governs the short-range quark-quark interaction in
the original chiral SU(3) quark model is now nearly replaced by
the vector-meson exchanges. We use these two models to study the
baryon-meson interactions.

In this paper, we show the RGM dynamical calculating results of
the $\Lambda K$, $\Sigma K$, $\Delta K$, $N\phi$, and $KN$ states
obtained in the chiral SU(3) quark model as well as in the
extended chiral SU(3) quark model
\cite{fhuang05lk,fhuang04dk,fhuang04nk,fhuang05nk,fhuang05nphi}.

\section{Formulation}

The chiral SU(3) quark model and the extended chiral SU(3) quark
model have been widely described in the literature
\cite{fhuang05lk,fhuang04dk,fhuang04nk,fhuang05nk}, and we refer
the reader to those works for details. Here we just give the
salient features of our chiral quark model.

In the chiral quark model, the total Hamiltonian of baryon-meson
systems can be written as
\begin{equation}
H=\sum_{i=1}^{5}T_{i}-T_{G}+\sum_{i<j=1}^{4}V_{ij}+\sum_{i=1}^{4}V_{i\bar
5},
\end{equation}
where $T_G$ is the kinetic energy operator for the center-of-mass
motion, and $V_{ij}$ and $V_{i\bar 5}$ represent the quark-quark
and quark-antiquark interactions, respectively,
\begin{equation}
V_{ij}= V^{OGE}_{ij} + V^{conf}_{ij} + V^{ch}_{ij},
\end{equation}
where $V^{ch}_{ij}$ is the chiral fields induced effective
quark-quark potential. In the chiral SU(3) quark model,
$V^{ch}_{ij}$ can be written as
\begin{equation}
V^{ch}_{ij} = \sum_{a=0}^8 V_{\sigma_a}({\bm r}_{ij})+\sum_{a=0}^8
V_{\pi_a}({\bm r}_{ij}),
\end{equation}
and in the extended chiral SU(3) quark model, $V^{ch}_{ij}$ can be
written as
\begin{eqnarray}
V^{ch}_{ij} = \sum_{a=0}^8 V_{\sigma_a}({\bm r}_{ij})+\sum_{a=0}^8
V_{\pi_a}({\bm r}_{ij})+\sum_{a=0}^8 V_{\rho_a}({\bm r}_{ij}).
\end{eqnarray}
$V_{i \bar 5}$ in Eq. (1) includes two parts: direct interaction
and annihilation parts,
\begin{equation}
V_{i\bar 5}=V^{dir}_{i\bar 5}+V^{ann}_{i\bar 5},
\end{equation}
with
\begin{equation}
V_{i\bar 5}^{dir}=V_{i\bar 5}^{conf}+V_{i\bar 5}^{OGE}+V_{i\bar
5}^{ch}.
\end{equation}
The annihilation interaction $V^{ann}_{i\bar 5}$ is not included
in baryon-meson interactions since they are assumed not to
contribute significantly to a molecular state or a scattering
process, which is the subject of our study.

\begin{table}[htb]
\tbl{Model parameters. $\Lambda=1100$ MeV, $m_u=313$ MeV,
$m_s=470$ MeV, $g_{ch}=2.621$, $b=0.5$ fm for set I and $0.45$ fm
for sets II and III. The meson masses are taken to be the
experimental data except for $m_\sigma$.}
{\begin{tabular}{@{}cccccccccc@{}} \toprule
    & $m_{\sigma}$ &$g_{chv}$&$f_{chv}/g_{chv}$& $g_u^2$ & $g_s^2$ & $a_{uu}^c$ & $a_{us}^c$ & $a_{ss}^c$ \\
    &     (MeV)    &         &                 &         &         &(MeV/fm$^2$)&(MeV/fm$^2$)&(MeV/fm$^2$)\\ \colrule
I   &      595     &   $-$   &       $-$       &  0.766  &  0.846  &    46.6    &    58.7    &    99.2    \\
II  &      535     &  2.351  &        0        &  0.056  &  0.203  &    44.5    &    79.6    &   163.7    \\
III &      547     &  1.973  &      $2/3$      &  0.132  &  0.250
& 39.1    &    69.2    &   142.5    \\ \botrule
\end{tabular}}
\end{table}

The model parameters can be fitted by several conditions, e.g. the
energies of baryons, the binding energy of deuteron, the chiral
symmetry, and the stability conditions of baryons. We listed them
in Table 1, where the first set is for the chiral SU(3) quark
model, the second and third sets are for the extended chiral SU(3)
quark model. All these three sets of parameters can give a
reasonable description of the $NN$ phase shifts \cite{lrdai03}.

\section{Results and Discussion}

\subsection{$\Lambda K$ and $\Sigma K$ States}

The nucleon resonance $S_{11}(1535)$ is explained as an excited
three quark state in the traditional constitute quark model
\cite{glozman96,isgur79}. But on the hadron level, it is explained
as a $\Lambda K$-$\Sigma K$ quasibound state
\cite{kaiser97,inoue02}. A dynamical study on a quark level of the
$\Lambda K$ and $\Sigma K$ interactions will be useful to get a
better understanding of the $S_{11}(1535)$ \cite{fhuang05lk}.

\vglue -0.2cm
\begin{figure}[h]
%\epsfxsize=10cm   %width of figure - will enlarge/reduce the figures
%\epsfbox{fig3.eps}
%\figurebox{2cm}{3cm}{} %to have a box alone
\centerline{\epsfxsize=4.6cm\epsfbox{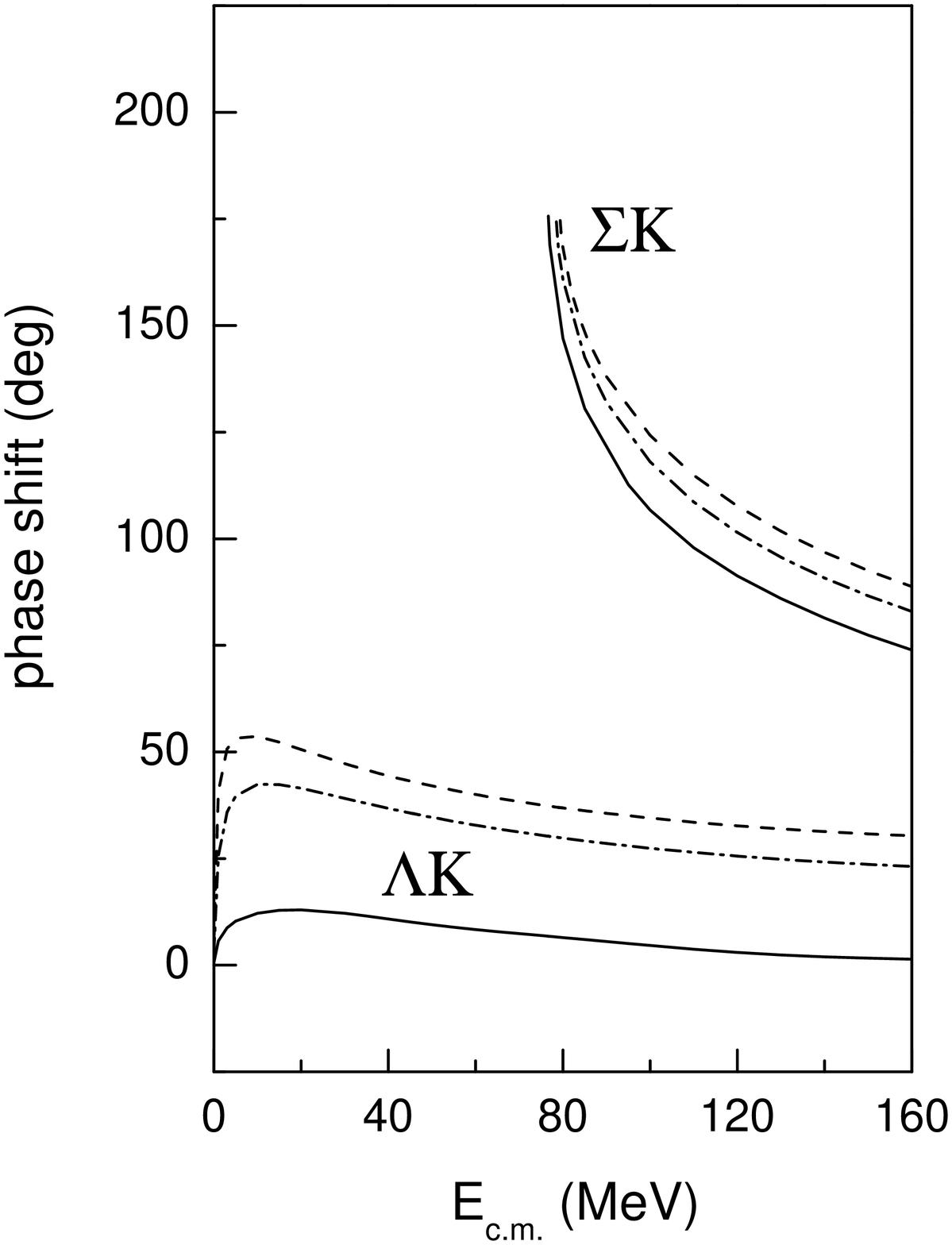}\epsfxsize=4.6cm\epsfbox{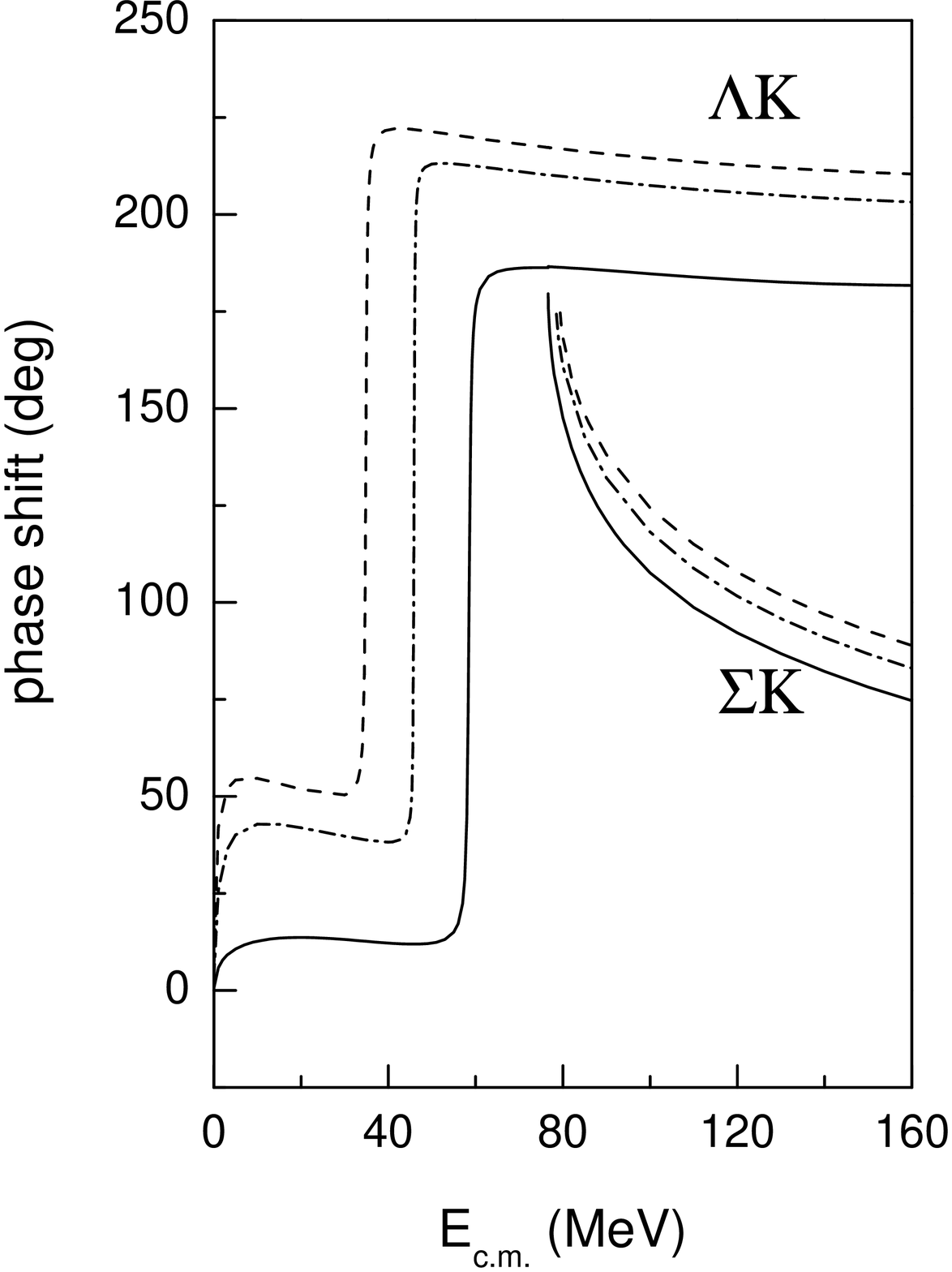}}
\vglue -0.4cm \caption{$S$-wave $\Lambda K$ and $\Sigma K$ phase
shifts in one-channel and coupled-channel calculation. The solid
curves represent the results obtained in the chiral SU(3) quark
model. The dashed and dash-dotted curves show the results from the
extended chiral SU(3) quark model by taking $f_{chv}/g_{chv}$ as
$0$ and $2/3$, respectively.}
\end{figure}

Fig. 1 shows the $S$-wave $\Lambda K$ and $\Sigma K$ phase shifts
in one-channel (left) and coupled-channel (right) calculations.
The phase shifts show that there is a strong attraction of $\Sigma
K$, which can make for a $\Sigma K$ quasibound state, while the
interaction of $\Lambda K$ is comparatively weak. When the channel
coupling of $\Lambda K$ and $\Sigma K$ is considered, the phase
shifts show a sharp resonance between the thresholds of these two
channels. The spin-parity is $J^P=1/2^-$ and width $\Gamma\approx
5$ MeV. The narrow gap of the $\Lambda K$ and $\Sigma K$
thresholds, the strong attraction between $\Sigma$ and $K$, and
the sizeable off-diagonal matrix elements of $\Lambda K$ and
$\Sigma K$ are responsible for the appearance of this resonance.
The final conclusion regarding what is the resonance we obtained
and its exact theoretical mass and width will wait for further
work where more channels will be considered.

\subsection{$\Delta K$ State}

The $\Delta K$ state has first been studied on the hadron level
\cite{sarkar05}. We perform a RGM dynamical study of the
structures of $\Delta K$ state within our chiral quark model
\cite{fhuang04dk}. Fig. 2 shows the diagonal matrix elements of
the Hamiltonian in the generator coordinate method (GCM)
calculation, which can describe the $\Delta K$ interaction
qualitatively.

\vglue 1.5cm
\begin{figure}[h]
%\epsfxsize=10cm   %width of figure - will enlarge/reduce the figures
%\epsfbox{fig3.eps}
%\figurebox{2cm}{3cm}{} %to have a box alone
\centerline{\epsfxsize=6.0cm\epsfbox{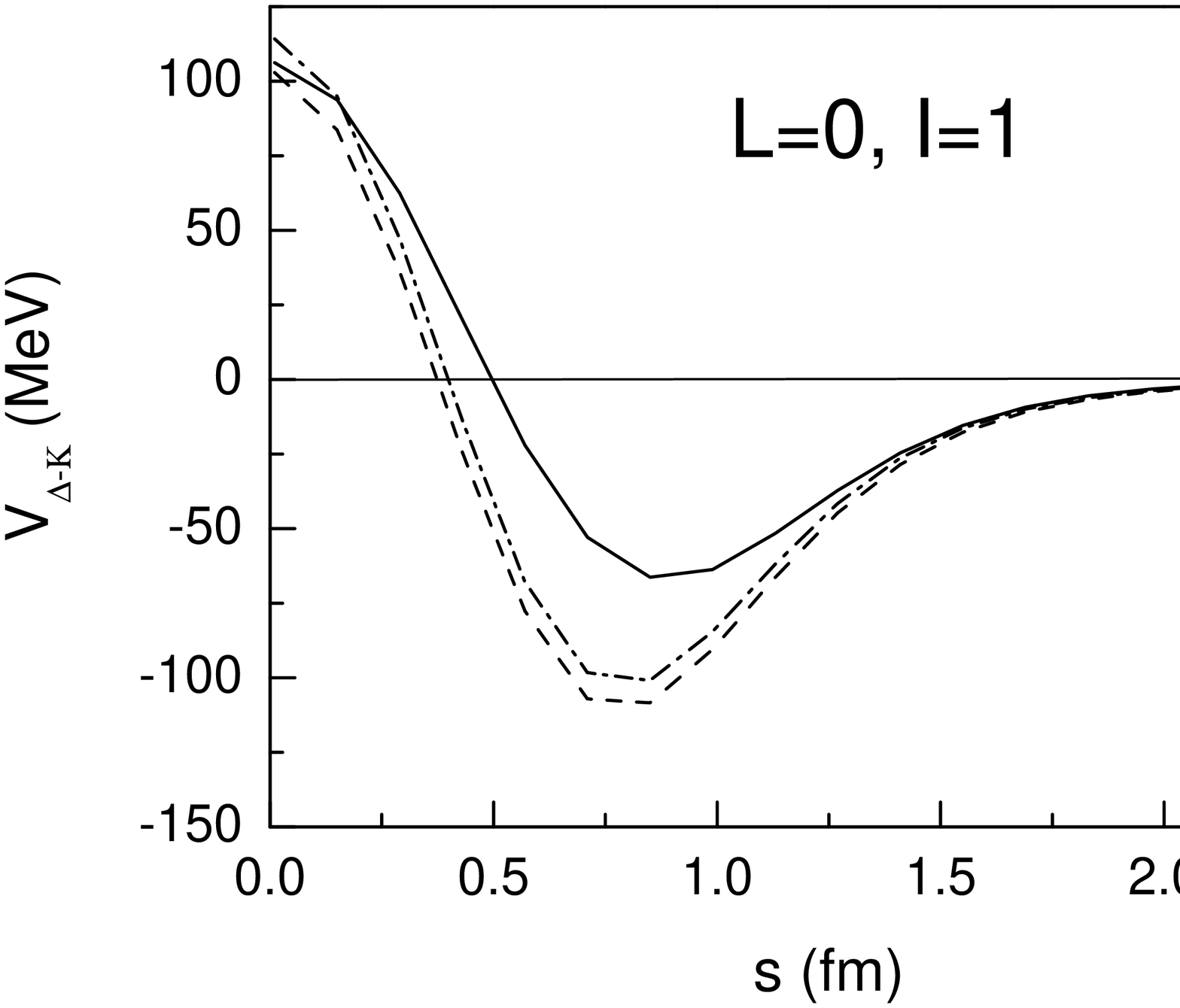}\epsfxsize=6.0cm\epsfbox{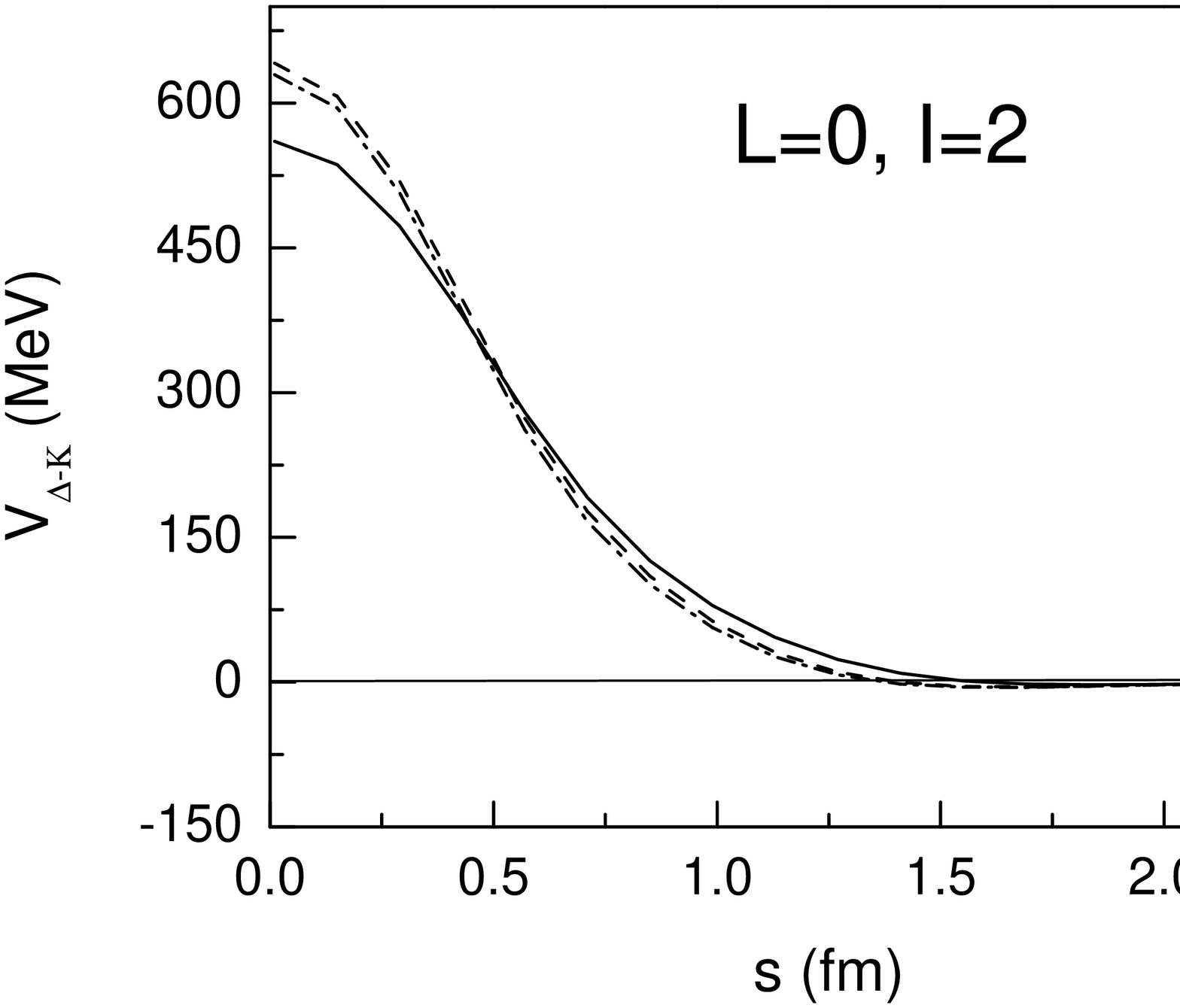}}
\vglue -1.8cm \caption{The GCM matrix elements of the Hamiltonian
for $\Delta K$. Same notation as in Fig. 1.}
\end{figure}

From Fig. 2 one sees that the $\Delta K$ interaction is attractive
for isospin $I=1$ channel, while strongly repulsive for $I=2$
channel. The attraction can result in a $\Delta K$ bound state
with a binding energy of about 3, 20, and 15 MeV by using three
different sets of parameters. To examine if ($\Delta
K$)$_{LSJ=0\frac{3}{2}\frac{3}{2}}$ is possible to be a resonance
or a bound state, the channel coupling between ($\Delta
K$)$_{LSJ=0\frac{3}{2}\frac{3}{2}}$ and
($NK^*$)$_{LSJ=0\frac{3}{2}\frac{3}{2}}$ would be considered in
our future work.

\subsection{$N\phi$ State}

It has been reported that the QCD van der Waals attractive
potential is strong enough to bind a $\phi$ meson onto a nucleon
to form a bound state \cite{gao01}. We try to study the
possibility of a $N\phi$ bound state in our chiral quark model
\cite{fhuang05nphi}. Table 2 shows the calculated binding energy
of $N\phi$.

\begin{table}[htb]
\tbl{The binding energy of $N\phi$.} {\footnotesize
\begin{tabular}{@{}ccccccc@{}}
\hline
 && & & & &\\[-1.6ex]
 && \multicolumn{2}{c}{One-channel} &&
\multicolumn{2}{c}{Coupled-channel} \\[0.8ex]  \cline{3-4} \cline{6-7}
\\[-1.6ex]
 && $S=1/2$ & $S=3/2$ && $S=1/2$ & $S=3/2$ \\[0.8ex]
\hline
 && && &&\\[-1.6ex]
I && $-$ & $-$& & $-$ & $-$ \\[0.8ex]
II && 1 & 3 && 3 & 9 \\[0.8ex]
III && $-$ & $-$ && 1 & 6 \\[0.8ex]
\hline
\end{tabular}}
\end{table}

From Table 2 one sees that in the one channel study, one can get a
bound state only by using the second set of parameters. When the
channel coupling to $\Lambda K^*$ is considered, the $N\phi$ is
found to be a bound state in the extended chiral SU(3) quark model
with several MeV binding energy. Further the tensor force will be
considered in the future work, which would make a bigger binding
energy.

For the $N\phi$ system, the two color-singlet clusters have no
quark in common. The attractive interaction is dominantly provided
by $\sigma$ exchange. Thus $N\phi$ is an ideal place to test the
strength of the coupling of the quark and $\sigma$ chiral field.

\subsection{$KN$ Scattering}

The $KN$ scattering has aroused particular interest in the past.
But most of the works on the quark level cannot give a reasonable
description of the $KN$ phase shifts up to $L=3$. We dynamically
study the $KN$ scattering in our chiral quark model
\cite{fhuang04nk,fhuang04dk,fhuang05nk}. When $m_{\sigma}$ is
chosen to be 675 MeV and the mixing of $\sigma_0$ and $\sigma_8$
is considered, we can get a satisfactory description of the $S$-,
$P$-, $D$-, and $F$-wave $KN$ phase shifts (see Figs. 1-4 in Ref.
4). The results from the chiral SU(3) quark model and the extended
chiral SU(3) quark model are quite similar although the
short-range interaction mechanisms in these two models are quite
different. Compared with the results in other's previous quark
model study \cite{lemaire03}, our theoretical phase shifts achieve
correct signs for several partial waves and a considerable
improvement in the magnitude for many channels.

\section{Summary}

We dynamically study the baryon-meson interactions in chiral quark
model by using the RGM. Some interesting results are obtained: (1)
The $\Sigma K$ state has an attractive interaction, which
consequently results in a $\Sigma K$ quasibound state. When the
channel coupling of $\Sigma K$ and $\Lambda K$ is considered, a
sharp resonance appears between the thresholds of these two
channels. (2) The interaction of $\Delta K$ state with $I=1$ is
attractive, which can make for a $\Delta K$ quasibound state,
while for $I=2$ channel, the interaction is strongly repulsive.
(3) When the coupling to the $\Lambda K^*$ channel is considered,
the $N\phi$ is found to be a quasibound state in the extended
chiral SU(3) quark model with several MeV binding energy. (4) The
calculated $KN$ phase shifts achieve a considerable improvement in
not only the signs but also the magnitudes in comparison with
other's previous quark model study.


\begin{thebibliography}{99}
\bibitem{fhuang05lk}F. Huang, D. Zhang, Z.Y. Zhang and Y.W. Yu, {\it Phys. Rev.} {\bf C71}, 064001 (2005).
\bibitem{fhuang04dk}F. Huang and Z.Y. Zhang, {\it Phys. Rev.} {\bf C70}, 064004 (2004).
\bibitem{fhuang04nk}F. Huang, Z.Y. Zhang and Y.W. Yu, {\it Phys. Rev.} {\bf C70}, 044004 (2004).
\bibitem{fhuang05nk}F. Huang and Z.Y. Zhang, {\it Phys. Rev.} {\bf C72}, 024003 (2005).
\bibitem{fhuang05nphi}F. Huang, Z.Y. Zhang, and Y.W. Yu, nucl-th/0512079.
\bibitem{lrdai03}L.R. Dai, Z.Y. Zhang, Y.W. Yu and P. Wang, {\it Nucl. Phys.} {\bf A727}, 321 (2003).
\bibitem{glozman96}L.Ya. Glozman and D.O. Riska, {\it Phys. Rept.} {\bf 268}, 263 (1996).
\bibitem{isgur79}N. Isgur and G. Karl, {\it Phys. Rev.} {\bf D19}, 2653 (1979).
\bibitem{kaiser97}N. Kaiser, T. Waas and W. Weise, {\it Nucl. Phys.} {\bf A612}, 297 (1997).
\bibitem{inoue02}T. Inoue, E. Oset and M.J. Vicente Vacas, {\it Phys. Rev.} {\bf C65}, 035204 (2002).
\bibitem{sarkar05}S. Sarkar, E. Oset and M.J.V. Vacas, {\it Eur. Phys. J.} {\bf A24}, 287 (2005).
\bibitem{gao01}H. Gao, T.S.H. Lee and V. Marinov, {\it Phys. Rev.} {\bf C63}, 022201 (2001).
\bibitem{lemaire03}S. Lemaire, J. Labarsouque and B. Silvestre-Brac, {\it Nucl. Phys.} {\bf A714}, 265 (2003).
\end{thebibliography}
\end{document}